\documentclass[%
 aip, cha,
 amsmath,amssymb,
preprint,%
]{revtex4-2}

\usepackage{graphicx}
\usepackage{dcolumn}
\usepackage{bm}

\usepackage[utf8]{inputenc}
\usepackage[T1]{fontenc}
\usepackage{mathptmx}
\usepackage{etoolbox}
\usepackage{nicefrac}
\usepackage{xcolor}
\usepackage{soul}
\usepackage{siunitx}
\usepackage{bm}

\usepackage{graphicx}
\usepackage{placeins}
\usepackage{booktabs}

\graphicspath{ {./Figures/} }

\DeclareMathOperator{\sech}{sech}

\makeatletter
\def\@email#1#2{%
 \endgroup
 \patchcmd{\titleblock@produce}
  {\frontmatter@RRAPformat}
  {\frontmatter@RRAPformat{\produce@RRAP{*#1\href{mailto:#2}{#2}}}\frontmatter@RRAPformat}
  {}{}
}%
\makeatother

\newcommand{\DDt}[1]{\frac{D #1}{Dt}}

\newcommand{\reva}[1]{\textcolor{black}{#1}}
\newcommand{\revb}[1]{\textcolor{black}{#1}}
\newcommand{\revboth}[1]{\textcolor{black}{#1}}

\usepackage{tikz}
\tikzset{every label/.style={font=\footnotesize,inner sep=1pt}}
\newcommand{\stencilptleft}[4][]{\node[circle,fill,draw,inner sep=1.5pt,label={left:#4},#1] at (#2) (#3) {}}
\newcommand{\stencilptcenter}[4][]{\node[circle,fill,draw,inner sep=1.5pt,label={above left:#4},#1] at (#2) (#3) {}}
\newcommand{\stencilptright}[4][]{\node[circle,fill,draw,inner sep=1.5pt,label={right:#4},#1] at (#2) (#3) {}}
\newcommand{\stencilptabove}[4][]{\node[circle,fill,draw,inner sep=1.5pt,label={above:#4},#1] at (#2) (#3) {}}
\newcommand{\stencilptbelow}[4][]{\node[circle,fill,draw,inner sep=1.5pt,label={below:#4},#1] at (#2) (#3) {}}

\begin{document}

\preprint{AIP/123-QED}

\title[Relevance of the Basset history term for Lagrangian particle dynamics]{Relevance of the Basset history term for Lagrangian particle dynamics}
\author{Julio Urizarna-Carasa}
\author{Daniel Ruprecht}%
 \affiliation{Lehrstuhl Computational Mathematics, Institut für Mathematik, Technische Universität Hamburg, 21073 Hamburg, Germany
}%

\author{Alexandra von Kameke}
\affiliation{%
Heinrich-Blasius-Institute, Faculty of Engineering and Computer Science, Hamburg University of Applied Sciences, 20099 Hamburg, Germany
}%

\author{Kathrin Padberg-Gehle}%
\email{kathrin.padberg-gehle@leuphana.de}
\affiliation{Applied Mathematics, Institute for Mathematics and its Didactics, Leuphana University L\"uneburg, 21335 L\"uneburg, Germany
}%

\date{\today}

%
\begin{abstract}
The movement of small but finite spherical particles in a fluid can be described by the Maxey-Riley equation (MRE) if they are too large to be considered passive tracers.
The MRE contains an integral "history term" modeling wake effects, which cause the force acting on a particle at some given time to depend on its full past trajectory.
The history term causes complications in the numerical solution of the MRE and is therefore often neglected, despite both numerical and experimental evidence that its effects are generally not negligible.
By numerically computing trajectories with and without the history term of a large number of particles in different flow fields, we investigate its impact on the large-scale Lagrangian dynamics of simulated particles.
We show that for moderate to large Stokes numbers, ignoring the history term leads to significant differences in clustering patterns.
Furthermore, we compute finite-time Lyapunov exponents and show that, even for small particles, the differences in the resulting scalar field when ignoring the BHT can be significant, in particular if the underlying flow is turbulent.
\end{abstract}
\maketitle
\begin{quotation}
Neglecting the Basset history term when solving the Maxey-Riley equations numerically can lead to significant changes not only of the simulated trajectories of individual particles but also of the resulting macroscopic Lagrangian dynamics.
\end{quotation}
%

\section{\label{sec:Intro}Introduction} %
Fluid motion is all around us, in the ocean, in the atmosphere, in industrial installations and in everyday live. 
Oftentimes, the fluid carries other materials with it, be it plankton, buoys, dust, particulate matter or any other type of small particulate pieces of immiscible material that does not dissolve in the carrier liquid. 
If the volume fraction of the particulate pieces is low, it can be assumed that they move in isolation and do not collide or affect each other's movement. 
Furthermore, if their mass fraction is low, the effect they have on the overall flow field of the carrier phase can be neglected.
Thorough discussions of the many different aspects of the mathematical description of the motion of inertial particles and real world applications can be found in the literature~\cite{marchioli_modeling_2017}. 

The equation of motion for inertial but small,  spherical, fully submerged particles that emerges from an inspection of the acting forces is called the Maxey-Riley equation (MRE)~\cite{maxey_riley_1983} but was also independently introduced by Gatignol~\cite{Gatignol1983}. 
The MRE accounts for the Basset history force, which models the influence of the particle's past accelerations on its present motion. 
The equation is valid for particles of intermediate size that are too large to be considered passive tracers but not so large that they significantly disturb the fluid or that surface effects become important.
The alternative is using a fully coupled fluid-structure interaction model, which, while detailed and realistic, requires massive computational effort and access to a powerful high-performance computing system~\cite{CostaEtAl2020}.

Mathematically, the MRE is a second order integro-differential equation. 
The integral or Basset history term (BHT), which captures the history force, is difficult to handle  numerically due to its non-local nature and is thus typically ignored~\cite{michaelides1992novel,farazmand_haller_2015}.
However, depending on the size and density of the particle, ignoring the history term changes the simulated trajectories significantly~\cite{Olivieri2014,DaitcheEtAl2014,prasath2019accurate}.
The importance of the BHT for matching simulations of particle trajectories to experiments has also been confirmed: Candelier et al.~\cite{CandelierEtAl2004} investigate the impact of the BHT experimentally to elucidate how a particle is ejected out of a vortex flow and find that ``calculations without history force overestimate particle ejection''. Similar observations have been made in simulations\cite{Guseva2013}.

In many applications, one is interested in the dynamics of ensembles of particles rather than in single particle trajectories. 
Such a macroscopic view on Lagrangian particle dynamics is particularly relevant in the context of studying transport and mixing processes. 
Lagrangian coherent structures (LCS) form the time-dependent skeleton of the flow, encompassing regions of stretching and folding that enhance or mitigate particle transport\cite{haller_2015}. 
Many different computational approaches have been developed over the past almost three decades to identify coherent flow structures, such as LCS or coherent sets, and to study their dynamics, including bifurcations~\cite{BenczikEtAl2002,Hadjighasem2017,Badza2023}.
Finite-time Lyapunov exponents (FTLE) are often heuristically used to highlight regions of different dynamical behavior. 
Inertial particles are known to interact with the underlying flow skeleton and tend to concentrate along different coherent structures, for example vortices, depending on their material properties such as size and density\cite{Sapsis2010,Sudharsan2016}. There are several studies that compare the FTLE fields for ideal tracers that exactly follow the underlying flow with those computed for inertial particles.
They found that the Lagrangian flow structures can be crucially different~\cite{garaboa2015method,Guenther2017}. In case of open chaotic flows, inertial tracer dynamics can have a strong impact on the underlying chaotic motion, which sensitively depends on the particles' properties\cite{BenczikEtAl2002}.  However, none of these studies considered the history term.
Therefore, this paper investigates the differences in flow structures for particle dynamics simulated with and without Basset term. For this, we make use of recent mathematical developments in the numerical solution of the Maxey-Riley equation with history term\cite{VanHinsbergEtAl2011,Daitche2015,prasath2019accurate,urizarna2024efficient}, which allow for the simulation of many particles~\cite{Haller2019}.  
An investigation similar to ours was carried out by Daitche et al.~\cite{DaitcheEtAl2011} for the von K\'{a}rm\'{a}n flow field. They analyze averaged Lyapunov coefficients and not the full FTLE field but reached very similar conclusions.
\subsection{\label{sec:Contributions}Contributions}
Our paper investigates the impact of neglecting the BHT on the macroscopic dynamics of simulated Lagrangian particles. 
We retain the assumption that the flow influences the particle motion but that there is no impact on the background flow, allowing us to use the MRE as model. 
We study the resulting flow patterns in a controlled setting for three example systems, the double gyre~\cite{shadden2005definition}, the Bickley jet~\cite{rypina2007lagrangian}, and a Faraday flow, a two-dimensional, fully turbulent, experimentally measured flow~\cite{colombi2021three,colombi2022coexistence}. The relatively large Stokes numbers considered in this work are inspired by the technological development of so-called Lagrangian sensors \cite{HofmannEtAl2022,HofmannEtAl2024} for reactor surveillance which we aim to describe by a type of Maxey-Riley equation in the future.

We compare particle dispersion and finite-time Lyapunov exponents (FTLE) for particle trajectories computed with and without history term for different Stokes numbers and densities. 
Our analysis demonstrates that, even for moderate Stokes numbers of unity or more, the  particle dispersion patterns revealed by the FTLE are noticeably changed by ignoring the history term. 
In particular, our results suggest that simulations without BHT produce results that are comparable to \revb{particles} with an effective Stokes number that is larger than the used value. 
This is very much in accordance with previous findings~\cite{CandelierEtAl2004,DaitcheEtAl2011,Guseva2013}. 
Our results also suggest that care must be taken when attempting to predict flow regimes, or sudden changes thereof, that involve inertial particle dynamics.
Important examples would be the oil spill in the Gulf of Mexico, ocean search-and-rescue~\cite{BeronVeraEtAl2019}, or the splitting of the polar vortex.

\subsection{Outline}\label{sec:Outline}
Section~\ref{sec:Background} introduces the the Maxey-Riley equation, its numerical solution and the computation of finite-time Lyapunov exponents. 
Section~\ref{sec:Fields} describes the three flow fields used in our study. 
In Section~\ref{sec:Results} we present the results of our investigation. 
First we analyse and compare the final particle positions and their preferential concentration.
Second, we present FTLE fields for a range of different parameter configurations. 
Conclusions and a summary can be found in Section~\ref{sec:conclusion}.

\section{Background}\label{sec:Background}%

In this section, we introduce the Maxey-Riley equation (MRE) and its numerical treatment as well as the equation and approximations used to obtain the finite-time Lyapunov exponents (FTLE).

\subsection{The Maxey-Riley Equation (MRE)}\label{sec:MRE}
The basic assumptions guaranteeing the validity of the  Maxey-Riley Equation used here are (i) spherical particles with a small radius compared to the flow dimensions, which guarantees a small particle Reynolds number~\cite{michaelides1992novel}, (ii) one-way interactions between particles and fluid, so that the flow influences the particle dynamics but not the other way round~\cite{prasath2019accurate} and, (iii) no particle collisions.
Under these assumptions, particle trajectories can be modeled using the MRE in nondimensionalized form
\begin{subequations}
    \label{eq:MRE}
    \begin{align}
        \label{eq:MRE:1}
        \bm{\dot{v}} =& \; \frac{1}{R} \DDt{\bm{u}} \; + \\
        \label{eq:MRE:2}
        &\; - \frac{1}{RS} \left( \bm{v} - \bm{u} \right) \; + \\
        \label{eq:MRE:3}
        &\; - \frac{1}{R} \sqrt{\frac{3}{\pi S}} \left\{ \frac{1}{\sqrt{t - t_0}} \left( \bm{v}(t_0) - \bm{u}(t_0) \right) + \int_{t_0}^t \frac{\bm{v}(s)-\bm{u}(s)}{\sqrt{t-s}}ds \right\},
    \end{align}
\end{subequations}
as adapted by Prasath et al.~\cite{prasath2019accurate} from the original paper by Maxey and Riley~\cite{maxey_riley_1983} with modifications by Auton et al.~\cite{AutonEtAl1988}, where $\bm{v} := \bm{\dot{x}}(t)$ is the particle's absolute velocity, $\bm{x}(t)$ the particle's position and $\bm{u}:=\bm{u}(\bm{x}(t),t)$ the value of the Eulerian velocity field at the particle's position.
All variables, such as time ($t$), space ($\bm{x}$) and velocities ($\bm{u}$ and $\bm{v}$) in equation~\eqref{eq:MRE} are nondimensional and obtained by dividing their dimensional value by the characteristic scales of the flow field.
Note that we only consider particles moving in a plane so that $\bm{x}(t), \bm{v}(t) \in \mathbb{R}^2$ throughout.
Further,
\begin{align}
    \beta := \frac{\rho_p}{\rho_f}, & & R := \frac{1+2\beta}{3}, & & S := \frac{1}{3}\frac{a^2}{\nu T},
\end{align}
where $\rho_p$, $\rho_f$ correspond to the particle and fluid densities, $a$ is the particle's radius, $\nu$ the fluids' kinematic viscosity and $T$ the time scale of the flow.
The definitions above are taken from the paper by Prasath et al.~\cite{prasath2019accurate} and are implemented like this in the accompanying code. However, it seems these definitions are not standardized and other authors use different expressions~\cite{farazmand_haller_2015}.
The effective density ratio $R$ controls buoyancy and inertia and accounts for the added mass effect that arises because the particle accelerates surrounding fluid and carries it along its path. 
Heavier particles that are denser than the carrier fluid with $\beta > 1$, where also $R>1$, are less susceptible to the fluid forces than lighter particles, i.e. $\beta < 1$ with $ R<1$, compare for the particle acceleration and first term in~\eqref{eq:MRE:1}.

The nondimensional parameter $S$ in~\eqref{eq:MRE} characterizes the particle radius with respect to the flow dimensions.
We refer to $S$ as the Stokes number since it compares the particle relaxation timescale $\tau_p = \frac{a^2}{\nu}$ to a flow timescale $T$. Our choices of S and R are in line with the paper that inspired this research \cite{prasath2019accurate}. 
Often, $T$ is chosen to be the time scale of the mean flow or eddy turnover time but setting it properly requires some care, since the timescales of the flow structures experienced by the particles should be considered~\cite{Huilier2021}.
We compute FTLE fields for values of $S=0.1$, $S=1$ and \revb{$S=3$}, even though the Maxey-Riley equation is strictly only valid for a Stokes number much smaller than one.
Our Stokes numbers are also slightly larger than values studied in other papers considering the Maxey-Riley Equation~\cite{Daitche2015,Olivieri2014} but roughly match Stokes numbers considered experimentally \cite{HofmannEtAl2022,Bourgoin2017,ZadeEtAl2019,ObligadoEtAl2014}.
For our density ratios of $\beta = \rho_p/\rho_f = 2/3$ and $\beta = 4/3$, a value of $S_{\textrm{exp}} = 1$ as studied by Olivieri et al.~\cite{Olivieri2014} corresponds to a Stokes number as used in this article of $S = 2.25$ and $S = 1.125$ respectively, since Olivieri et al. use the experimental Stokes number as introduced below. The largest Stokes number studied by Daitche\cite{Daitche2015} is $S_{\textrm{Daitche}} = 3$, corresponding to $S = 3.86\, (\beta = 2/3)$ and $S = 2.46\, (\beta = 4/3)$, where $S_{\textrm{Daitche}} = S \cdot R$ and thus taking into account the added mass term in his definition of the Stokes number.

The Stokes numbers in experimental references \cite{OuelletteEtAl2008, Olivieri2014} often corresponds to  
\begin{equation}
    S_{exp} = \frac{\tau_p}{T}  =  S \cdot \frac{2}{3} \frac{\rho_p}{\rho_f}= \frac{2}{9} \frac{a^2}{\nu}\frac{\rho_p}{\rho_f} \frac{1}{T} = \frac{2}{9} \frac{a^2}{L^2}\frac{\rho_p}{\rho_f} \frac{U L}{\nu} = \frac{2}{9} \frac{a^2}{L^2}\frac{\rho_p}{\rho_f} Re_f 
\end{equation}
with $T$ being the flow time scale calculated by a typical flow velocity and length scale $U$, $L$ which can be reordered to give a flow Reynolds number $Re_f$.
The experimental Stokes number is slightly different from the nondimensional $S$ we prescribe here. 
It arises as a prefactor of the Stokes drag term, formulated without the added mass term, in the nondimensionalized Maxey-Riley equation and is widely used in the literature~\cite{OuelletteEtAl2008, HofmannEtAl2022}. 
It allows to account for non-Stokesian drag when $Re > 1$ by using the expression $S_{\textrm{exp}} = f \cdot \frac{a^2}{\nu}\frac{\rho_p}{\rho_f} \frac{1}{T}$, where $f$ is a factor that depends on the particle Reynolds number $f = f(Re_p)$.~\cite{HofmannEtAl2022}

The MRE is derived from Newton's second law for particle movement and a velocity field by approximating the influence of the particle on the local flow from potential flow theory under creeping flow assumption. 
This implies that the validity of the model is only guaranteed for low particle Reynolds numbers with low relative velocities.
Terms on the right hand side of equation~\eqref{eq:MRE} correspond to different forces acting on the particle.
Term~\eqref{eq:MRE:1} are the forces acting on a particle in an unperturbed fluid lumped up with the added mass effect.
Term~\eqref{eq:MRE:2} is the Stokes drag.
Term~\eqref{eq:MRE:3}, the so-called Basset History Term (BHT), accounts for the effects of the lagging boundary layer around the sphere~\cite{langlois2015asymptotic}.

Since the aim of our work is to examine the influence of this history term (BHT), we neglect the so-called Faxen corrections that model the influence of the particle's finite size and scale with the curvature of the velocity field. 
However, for larger particles, the Faxen corrections should be considered as they have been shown to play a crucial role in shear flows~\cite{ChongEtAl2013} where they cause fluctuations in the particle's velocity and thus a relative velocity between particle and fluid flow. 
In~\eqref{eq:MRE}, since we assume that the particle initially has same velocity as the fluid, only particles with buoyancy $R \neq 1$ will produce non-zero relative velocities and cause the trajectories of particles with and without history force to deviate. It is further important to note that the particles are assumed to be fully immersed in the liquid. 
In the following, particle behavior in two-dimensional flows is considered, and thus the particles are better visualized as circles rather than spheres. 
Effects of the interface of the liquid and a gaseous phase velocity onto the movement as described elsewhere~\cite{BeronVeraEtAl2019} are not considered.
For Stokes numbers $S \in [0.1, 1.0]$, the mean particle Reynolds numbers in the Faraday flow are of order unity with maximum values of $Re_p \approx 10$\revb{based on the maximal slip velocity along a particle trajectory}. 
There, it can be assumed that the Maxey-Riley equations are a valid approximation based on former experimental analyses with particle Reynolds numbers up to $Re_p \approx 17$ (as in refs.~\onlinecite{DaitcheEtAl2014,MaxeyEtAl1996}) and $Re_p = 2.5$ (ref.~\onlinecite{Abbad2004}). 
However, for \revb{$S=3$} and the Faraday flow, we note that the underlying assumptions for the derivation of the Maxey-Riley equation are partly not fulfilled. \revb{
In this case, the Reynolds number of the particle\cite{DaitcheEtAl2014} $Re_p = \frac{a \cdot u_{slip_{max}}}{\nu}$, when calculated with the maximum slip velocity values along a particle trajectory, can reach values of $Re_p \approx 37$, and, based on the mean slip velocity along the particle trajectory, values of $Re_p \approx 5$}. 
For \revb{$Re_p \approx 37$}, there is currently no experimental or analytic evidence for the validity of the approximation. 
Nevertheless, since we aim to eventually use a type of Maxey-Riley equation to model Lagrangian sensors, we still analyze the impact of neglecting the BHT for \revb{$S=3$}.

\subsection{Numerical solution of the MRE}
The BHT is made up of a history integral with a singular kernel.
This term changes the MRE from an ordinary differential into an integro-differential equation that is not easily solvable either analytically or numerically.
For this reason, the term is often omitted~\cite{CumminsEtAl2020}, modified~\cite{LovalentiEtAl1993, Mei1994, DorganAndLoth2007, MorenoCasasEtAl2016} or approximated~\cite{KlinkenbergEtAl2014, ElghannayAndTafti2016, ParmarEtAl2018, VanHinsbergEtAl2011}.
Several numerical approximations, based on quadrature schemes, were obtained by Van Hinsberg et al.~\cite{VanHinsbergEtAl2011} and Daitche~\cite{Daitche2013}.
However these schemes become storage-intensive for large time grids.

In this paper, trajectories with BHT are calculated with the second order finite-difference, IMEX solver (FD2 + IMEX2) by Urizarna-Carasa et al.~\cite{urizarna2024efficient}.
The scheme uses the reformulation by Prasath et al.~\cite{prasath2019accurate} that transforms the MRE into a ``a forced, time-dependent Robin boundary condition of the one-dimensional diffusion equation" on a semi-infinite domain
\begin{subequations}
\label{eq:MRSys}
\begin{align}
		\bm{q}_t(z,t) &= \bm{q}_{zz}(z,t), & z>0, t\in(0,T], \label{MRSys1} \\
		\bm{q}(z,0) &= \bm{0}, & z>0, \label{MRSys2}\\
		\bm{q}_t(0,t) + \alpha\bm{q}(0,t) - \gamma\bm{q}_z(0,t) &= \bm{f}(\bm{q}(0,t),\bm{x}(t),t), & t\in[0,T], \label{MRSys3} \\
		\dot{\bm{x}}(t) &= \bm{q}(0,t) + \bm{u}(\bm{x}(t),t), & t\in[0,T], \label{MRSys4} \\
		\lim_{t\to 0}\bm{q}(0,t) &= \bm{v}_0 - \bm{u}_0, \label{MRSys5}\\
		\bm{x}(t_0) &= \bm{x}_0. \label{MRSys6}
\end{align}
\end{subequations}
\revb{The transformation relies on the observation that the integral term in~\eqref{eq:MRE} is equivalent to a fractional derivative of Riemann-Liouville type~\cite{Tatom1988}.
For a detailed discussion of how~\eqref{eq:MRE} is transformed into~\eqref{eq:MRSys} we refer the reader to the paper by Prasath et al.~\cite{prasath2019accurate}.}
Here, $z \in \mathbb{R}^+$ is a pseudo-space that has no physical interpretation, $t$ is time, $\bm{x}(t)$ is the particle's position at time $t$ and $\bm{q}(z,t)$ is a function with the same dimension as $\bm{x}(t)$, that is two for a two-dimensional flow field and three for a 3D field.
The boundary value $\bm{q}(0,t)$ is equal to the relative velocity of the particle at time $t$ and 
$\dot{\bm{x}}(t)$ is its absolute velocity.
\revb{The parameters in the boundary condition~\eqref{MRSys3} are
\begin{equation}
    \alpha = \frac{1}{RS} \quad \text{and} \quad \gamma = \frac{1}{R} \sqrt{\frac{3}{S}}.
\end{equation}
The function $\bm{f}$ is given by the flow field
\begin{equation}
    \bm{f}\left( \bm{q}(0,t),\bm{y}(t),t \right) = \left( \frac{1}{R} - 1 \right) \DDt{u} - \bm{q}(0,t) \cdot \nabla_y \bm{u}(\bm{y}(t),t).    
\end{equation}
}

The FD2 + IMEX2 scheme we employ to solve~\eqref{eq:MRE} uses the second order spatial discretization provided by Koleva~\cite{Koleva2005} to cope with the semi-infinite domain to discretize in space.
The resulting semi-discrete initial value problem is then integrated with the second-order implicit-explicit midpoint rule by Ascher et al.~\cite{ascher1997implicit} to avoid having to use a costly Newton solver for the nonlinear boundary condition.

Trajectories without BHT are calculated with the explicit adaptive Runge-Kutta method of order 5(4) provided by the \textit{solve\_ivp} solver of Python's SciPy~\cite{2020SciPy-NMeth} library, using a relative and absolute tolerance of $10^{-8}$.
For verification, trajectories without BHT were compared against the Leap-Frog method, whereas trajectories with BHT were compared against Daitche's $3^{rd}$ order method~\cite{Daitche2013}.
All figures shown in this paper can be reproduced using the provided code~\cite{code2024} available at \url{https://doi.org/10.5281/zenodo.14252124}.

\subsection{Finite-time Lyapunov exponents (FTLE)}\label{sec:FTLE_background}
Finite-time Lyapunov exponents (FTLE) are a measure of the exponential growth of an infinitesimal perturbation in the initial condition under the action of the dynamical system. 
FTLE fields have become a popular, heuristic tool in nonlinear dynamics to highlight organizing structures in phase space such as invariant manifolds of hyperbolic objects or Lagrangian coherent structures in time-dependent flows\cite{shadden2005definition,haller_2015}.  Typically FTLE have large values for repelling LCS when computed in forward time, and for attracting ones when computed in backward time. Close to zero values are frequently observed for elliptic motion, such as in the center of vortices. 
While more sophisticated approaches have been developed to identify LCS~\cite{haller_2015}, we use FTLE as a convenient and meaningful scalar field that allows us to visualize regions of different dynamical behavior. 

The expression to compute the FTLE of a flow field over the time interval $[t_0, t_{end}]$ at a position $\bm{x}$ is
\begin{eqnarray}
    \label{eq:FTLE}
    \sigma_{t_0}^{t_{end}} (\bm{x}) := \frac{1}{\lvert t_{end}-t_0 \rvert}\ln \left( \sqrt{ \lambda_{max}(\Delta_{t_0}^{t_{end}})} \right),
\end{eqnarray}
where $\bm{x}=(x, y)^T$ is the position vector at time $t_0$, $\lambda_{max}(\Delta)$ is the largest eigenvalue of the Cauchy-Green deformation tensor $\Delta_{t_0}^{t_{end}} := \left(D\Phi_{t_0}^{t_{end}}\right)^*D\Phi_{t_0}^{t_{end}}$, and $D\Phi_{t_0}^{t_{end}}$ is the deformation gradient of the flow map $\Phi_{t_0}^{t_{end}}$, which maps particles initialized at time $t_0$ to their position at time $t_{end}$. 
The choices for $t_0$ and $t_{end}$ will be discussed when introducing the different flow fields in Section~\ref{sec:Fields}.
The flow map $\Phi_{t_0}^{t_{end}}$ is evaluated by integrating~\eqref{eq:MRE} using either \textit{solve\_ivp} or the FD2 + IMEX2 scheme described above.
$D\Phi_{t_0}^{t_{end}}$ can either be obtained by solving a variational equation or by approximating the derivative numerically using finite-differences. The latter is straightforward to implement and thus used in most cases, even when a kinematic model is available. 

FTLE methods have also been applied in the context of inertial particle dynamics described by the Maxey-Riley equation~\cite{garaboa2015method,Sudharsan2016,Guenther2017} and perturbations both in positions and velocities have been taken into account to define inertial finite-time Lyapunov exponents (iFTLE). In that case, not only partial derivatives with respect to positions but also to velocities would make up the deformation gradient of the flow map, resulting in a $4\times 4$ matrix in the case of a two-dimensional flow velocity field.
Since our aim is to study the influence of the Basset history term on particle dynamics and transport properties by comparing flow structures computed with and without history term we neglect perturbations in the initial velocities and the flow map only considers the particle positions.
Therefore, in our study, $D\Phi_{t_0}^{t_{end}}$ is a $2\times 2$ deformation gradient. 

Numerical particle trajectories can only be calculated at discrete points in space. 
Thus, the entries of the deformation gradient of a particle in an interior point of the domain (position $\bm{x}_{i,j}=(x_{i,j}, y_{i,j})$ in the finite difference stencil shown in Figure~\ref{fig:stencil}) are approximated by the second order finite-difference scheme~\cite{garaboa2015method}
\begin{subequations}
\begin{eqnarray}
    D\Phi_{t_0}^{t_{end}} \rvert_{1,1} &\approx \frac{x_{i+1,j}(t_{end}) - x_{i-1,j}(t_{end})}{x_{i+1,j}(t_0) - x_{i-1,j}(t_0)}, \\
    D\Phi_{t_0}^{t_{end}} \rvert_{1,2} &\approx \frac{x_{i,j+1}(t_{end}) - x_{i,j-1}(t_{end})}{y_{i,j+1}(t_0) - y_{i,j-1}(t_0)}, \\
    D\Phi_{t_0}^{t_{end}} \rvert_{2,1} &\approx \frac{y_{i+1,j}(t_{end}) - y_{i-1,j}(t_{end})}{x_{i+1,j}(t_0) - x_{i-1,j}(t_0)}, \\
    D\Phi_{t_0}^{t_{end}} \rvert_{2,2} &\approx \frac{y_{i,j+1}(t_{end}) - y_{i,j-1}(t_{end})}{y_{i,j+1}(t_0) - y_{i,j-1}(t_0)}.
\end{eqnarray}
\end{subequations}
At the boundaries, where values outside the initial domain are not available, a first order approximation is used.
Following Shadden et al.~\cite{shadden2005definition}, all FTLE plots in this paper show $\sigma_{t_0}^{t_{end}}(\bm{x})\lvert t_{end}-t_0\rvert$, hence a nondimensional value.
\begin{figure*}[th]
\centering
\begin{tikzpicture}
  \node (a) at (-4,0)
  {
  \begin{tikzpicture}
  \stencilptleft[red]{-1,0}{i-1}{$\bm{x}_{i-1,j}(t_0)$};
  \stencilptcenter{ 0,0}{i}  {$\bm{x}_{i,j}$}; 
  \stencilptright[red]{ 1,0}{i+1}{$\bm{x}_{i+1,j}(t_0)$};
  \stencilptbelow[blue]{0,-1}{j-1}{$\bm{x}_{i,j-1}(t_0)$};
  \stencilptabove[blue]{0, 1}{j+1}{$\bm{x}_{i,j+1}(t_0)$};
  \draw (j-1) -- (i)
        (i)   -- (j+1)
        (i-1) -- (i)
        (i)   -- (i+1);
  \end{tikzpicture}};
  \node (b) at (4,0) 
  {
  \begin{tikzpicture}
  \stencilptleft[red]{-0.75,0}{i-1}{$\bm{x}_{i-1,j}(t_0 + \Delta t)$};
  \stencilptcenter{ 0,0}{i}  {$\bm{x}_{i,j}$}; 
  \stencilptright[red]{ 0.75,0}{i+1}{$\bm{x}_{i+1,j}(t_0 + \Delta t)$};
  \stencilptbelow[blue]{-1,-1}{j-1}{$\bm{x}_{i,j-1}(t_0 + \Delta t)$};
  \stencilptabove[blue]{1, 1}{j+1}{$\bm{x}_{i,j+1}(t_0 + \Delta t)$};
  \draw (j-1) -- (i)
        (i)   -- (j+1)
        (i-1) -- (i)
        (i)   -- (i+1);
  \end{tikzpicture}};
  \draw[->,thick] (a) -- (b) node [pos=0.5,above] {$ \Phi_{t_0}^{t_0+\Delta t}$};
  \draw [->] (a) -- (b);
\end{tikzpicture}
\caption{\label{fig:stencil} Finite difference stencil used for the calculation of the FTLE $\sigma_{t_0}^{t_{end}}(\bm{x}_{i,j})$ at position $\bm{x}_{i,j}$, illustrated as an example of the translation of the particles in the stencil after a small time $\Delta t$.}
\end{figure*}

\section{Flow fields}\label{sec:Fields}
This section introduces the three flow fields that will serve as test cases for our study. 
Our choice of two-dimensional velocity fields includes the frequently studied kinematic double gyre and Bickley jet as well as a Faraday flow obtained from experimental measurements.
In all three fields, particle trajectories are numerically integrated from an initial time $t_0$ until a final time $t_{end}$ with a timestep of $\Delta t=0.01$. The choices of $t_0$ and $t_{end}$ are stated below.
Particles always start with zero relative velocity, that is $\bm{v}(t_0) = \bm{u}(\bm{x}(t_0),t_0)$.

\subsubsection{Double gyre}\label{sec:Gyre}

The double gyre is a popular example in the study of Lagrangian coherent structures and FTLE.
It is a simplified model of a pattern that occurs in geophysical flows~\cite{shadden2005definition}.
The velocity flow field is obtained from the  streamfunction 
\begin{align}
    \label{eq:DoubleGyre}
    \psi(x, y, t) = A \sin(\pi f(x,t)) \sin(\pi y),
\end{align}
where
\begin{subequations}
\begin{align}
    f(x,t) &= a(t) x^2 + b(t) x, \\
    a(t) &= \varepsilon \sin(\omega t), \\
    b(t) &= 1 - 2\varepsilon\sin(\omega t),
\end{align}
\end{subequations}
with $A=0.1$, $\varepsilon = 0.25$, $\omega=\frac{\pi}{5}$.
The double gyre is already in its nondimensional form with characteristic timescale $T_{DG}=\nicefrac{2\pi}{\omega} = \SI{10}{\second}$.
The characteristic length scale is $L_{DG} = \SI{1}{\meter}$.

We consider $20301$ particles starting in the rectangle $M = [0,2]\times[0,1]$, which is an invariant set for passive tracers and frequently used in the literature\cite{shadden2005definition}.
Particles are distributed in $201$ vertical by $101$ horizontal lines, so that adjacent particles are separated by a distance of $0.01$.
Unlike ideal passive tracers, inertial particles may leave $M$. 
As the streamfunction is defined on $\mathbb{R}^2$, particles that have left $M$ continue to be advected by the flow and thus their full-length trajectories can be used for FTLE computations. 
We note that due to periodic forcing, the velocity field on $M$ cannot be periodically continued in $x$. We consider the flow on the time interval $[t_0, t_{end}]$, with $t_0=0$ and $t_{end}=10$ in nondimensional units. Since particles may leave $M$ and never return, we have the setting of an open or leaking flow, with particles staying in $M$ for long or even all times potentially tracing out influential flow structures, such as invariant manifolds of a chaotic saddle\cite{Aref2017frontiers}.

\subsubsection{Bickley jet}\label{sec:Bickley}

As our second example, we consider the Bickley jet proposed by Rypina  et al. ~\cite{rypina2007lagrangian}. 
It is defined by the streamfunction
\begin{subequations}
\begin{align}
\Psi(x,y,t)  =&   -U_0 L \tanh\left(\frac{y}{L}
\right) + \\
&+\sum_{i=1}^3 A_i U_0 L\sech^2\left(\frac{y}{L}\right)\cos(k_i x-\sigma_i t), \label{eq:bickley}
\end{align}
\end{subequations}
and serves as an idealized model of stratospheric flow.
We use the same parameter values as Hadjighasem et al.~\cite{hadjighasem2016spectral, padberg2017network}, i.e.\  $U_0=5.414$, $A_1=0.0075$, $A_2=0.15$, $A_3=0.3$, $L=1.770$,  $c_1/U_0=0.1446$, $c_2/U_0=0.205$, $c_3/U_0=0.461$, $k_1=2/r_e$, $k_2=4/r_e$, $k_3=6/r_e$ where $r_e=6.371$ as well as $\sigma_i=c_ik_i$, $i=1,2,3$.
$U_0$ and $L$ correspond to the characteristic velocity and length of the field, measured in $\si{\mega\meter\per\day}$ and $\si{\mega\meter}$, respectively.
The characteristic timescale thus results in $T_{Bickley} = \nicefrac{L}{U_0} = \SI{0.327}{\day}$.

In our study, $16281$ initial conditions are chosen in the rectangle $M=[0, 20[\times [-4, 4]$.
Particles are distributed within $81$ horizontal lines by $201$ vertical lines, so that again there is a separation of $0.01$ between adjacent particles.
Periodic boundary conditions are imposed in the $x$-direction and thus the Eulerian velocity is considered on a cylinder. 
For this setting, the flow exhibits a meandering central jet and three regular vortices on each side of the jet~\cite{rypina2007lagrangian,padberg2017network}. Inertial particles may potentially leave $M$ in the vertical direction. 
Since the velocity field is available outside of $M$ (i.e. also for $|y|>4$), we can use full-length trajectories again for FTLE computations. 
As in a previous study~\cite{padberg2017network}, we consider the flow over the time interval $[t_0, t_{end}]$ with $t_0=10$ and $t_{end}=30$ in nondimensional units.

\subsubsection{Faraday flow}\label{sec:Faraday}
The Faraday flow field is an experimentally-measured flow field. 
The data was recorded in the course of the master thesis of J. Tenhaus under the supervision of A. v. Kameke.
The velocity fields are obtained from the surface of a \SI{2}{\mm}-thick layer of distilled water in a cylindrical container of \SI{290}{\mm} diameter as it is vertically shaken with a monochromatic sinusoidal signal at a frequency of \SI{50}{\hertz} at a measured forcing acceleration of $a = 1.6 g$ which corresponds to a supercriticality of $\epsilon = 0.04$~\cite{colombi2021three}.  
This movement produces the so-called Faraday waves, quasi-standing waves over the fluid's surface, and a turbulent space- and time-dependent 2D-velocity field which exhibits all characteristics of two-dimensional turbulence~\cite{vonKamekeEtAl2011}. 
Energy is injected into the flow at a scale of half the Faraday wavelength which corresponds to the spatial scale of the smallest occurring vortices. 
In contrast to three-dimensional turbulence, energy is passed from this scale upwards to larger scales and an inverse energy cascade is forming, distributing the energy from the forcing scale up to the system size. 
Therefore, above half the Faraday wavelength all vortex sizes occur, leading to effective turbulent diffusion of tracers in the flow above the forcing scale~\cite{vonKamekeEtAl2013}.
The velocity field is measured using particle image velocimetry (PIV) with floating white, hollow, glass microspheres (diameter of $70 \mu m$, density of $0.15 g/cm^3$, Fibre Glast) to visualize the horizontal motion. 
The tracer particles are chosen so that they follow the fluid motion as closely as possible. 
An upper estimate for the experimental Stokes number of the particles in the Faraday flow is  $S_{\textrm{exp}}^{\textrm{max}} = 0.0020$ using the maximal flow velocity and the half the Faraday wavelength as typical velocity and length scales. 
This is justified since the smallest scales in the Faraday flow are vortices of $\lambda_f/2 = 5\, mm$.
This corresponds to a Stokes number $S = 0.0031$, which is nearly two orders of magnitude smaller than the smallest Stokes number $S = 0.1$ considered here.
A non-ionic surfactant ($1 \% $ polysorbate 80) is used in a solution of $10 \% $ or less to ensure that the particles do not aggregate and sink. 
The particles are recorded with a high-speed camera (pco.dimax HS2 and Carl Zeiss Makro-Planar $T* 2.8/100mm$ lens) and triggered at a frequency of \SI{400}{\hertz}.
The details of the experimental setup are similar to those given by Colombi et al.~\cite{colombi2022coexistence, colombi2021three}. 
From the particle images, velocity data are obtained using PIVview (PIVTEC GmbH, Germany) and
MATLAB. 
One set of data for the Faraday flow consists of discrete velocity values for both horizontal and vertical velocity components, $vx$ and $vy$, at $115 \times 86$ spatial grid points over an area of $70.395 \times 52.487\, mm^2$ as results from a length conversion via calibration.
These measurements were taken for $1056$ successive time steps, resulting in a total measurement time of \SI{42.24}{\second}.

For the calculation of the FTLE in the Faraday flow, the measured velocity field is used to advect inertial particles numerically. 
It should be stressed that experimental particles are not further considered for analysis, they solely serve to derive the time-dependent velocity field.
The mean square velocity $v_{rms}= \SI{ 0.004862}{\meter\per\second}$ is taken as the characteristic velocity of the flow and the vertical length of the domain, $\SI{0.052487}{\meter}$ as the \revb{characteristic} length-scale.
A timescale is derived by dividing the characteristic length-scale by the characteristic velocity, obtaining $T_{Faraday} \approx \SI{10.795254}{\second}$.
Simulated numerical trajectories are computed for $40401$ particles initialized in the rectangle $M=[0,0.070395]\,m\times[0,0.052487]\,m$ in the experimentally derived velocity field. 
Simulated numerical particles of varying Stokes number as well as simulated perfect tracer particles are distributed along $201$ vertical and $201$ horizontal lines and are allowed to leave the initial domain, since no boundary condition is imposed.
The flow field's velocity and its derivatives are set to $0$ beyond the boundaries of $M$, so that particles behave as relaxing particles after they leave the initial domain.
They continue in a straight line until they reach a halt due to the friction with the fluid and thus are unable to return. 
By this trivial extension of the velocity field, we again obtain full-length trajectories for FTLE computations. The flow is studied on the time interval $[t_0, t_{end}]$ with $t_0=\SI{0}{\second}$ and $t_{end}=\SI{10}{\second}$, so only the first 10 seconds of the 
total \SI{42.24}{\second} are considered.
When needed, intermediate velocity values are interpolated in space with SciPy’s bivariate rectangular Spline~\cite{2020SciPy-NMeth}.
Interpolation in time is carried out using linear interpolation. 
\section{Numerical results}\label{sec:Results}

In this section, we analyze the impact of the history term first on the final distribution of particles and then on the FTLE.
The flow fields we consider are a double gyre, the Bickley jet and an experimentally measured Faraday flow as described in Section~\ref{sec:Fields}.

\subsection{Particle clustering}\label{sec:Clusters}
\revb{For all three flow fields,} we show the final positions of all particles that have not left the region $M$ they were initialized in, first for lighter-than-fluid and then for heavier-than-fluid particles.
\revb{Moreover}, we state the relative average difference in the final position of particles computed with and without BHT
\begin{equation}
    \label{eq:distance}
    d := \frac{1}{N} \sum_{i=1}^N \left( \frac{\left\| \bm{x}_i^{\text{History}}(t_{\text{end}}) - \bm{x}_i^{\text{Stokes}}(t_{\text{end}})\right\|_2}{\frac{1}{N} \sum_{j=1}^N \left\| \bm{x}_j^{\text{History}}(t_{\text{end}}) - \bm{x}_j^{\text{History}}(t_0) \right\|_2} \right).
\end{equation}
in Table~\ref{tab:diff}.
Here, $N$ is the number of particles, $\bm{x}_i^{\text{History}}(t_{\text{end}})$ the final position of a particle computed with the full MRE and $\bm{x}_i^{\text{Stokes}}(t_{\text{end}})$ the final position of the same particle computed without BHT.

The average relative difference in final positions computed with and without BHT for the double gyre and Bickley jet with $S=0.1$ are less than 10\% with standard deviations of similar order.
However, for the more turbulent Faraday flow that contains flow structures spanning a large range of scales, even for $S=0.1$ we see a significant effect from the BHT. 
This can likely be explained by the chaotic characteristics of the Faraday flow which, due to its turbulence in space and the large fluctuations in time, leads to a fairly dense tangle of Lagrangian coherent structures, where nearby particles get caught up in very different fates. 
Further, the lowest spatial scale in the Faraday flow, half the Faraday wavelength $\lambda_f /2 \approx 5 mm$ is smaller when compared to the simulated particle sizes. 
For instance calculating the size of the simulated particle \revb{for $S = 0.1, 1, 3$} via 
\begin{equation}
a_p = \sqrt{S \beta \frac{9}{2} \frac{(\lambda_f /2)^2}{Re_f}}
\end{equation}
results in radii \revb{$a_p \approx [500, 1700, 3000] \; \mu m$}  for $\beta = 4/3$. 
\revb{For our choices of $S = 0.1$ and $S = 1$}, the particle Reynolds number can be considered small enough, as detailed above, and the flow around the particles can be considered to be smooth since there are almost no velocity structures below half the Faraday wavelength $\lambda_f /2 \approx 5 \,mm$. 
\revb{This is, however, debatable for the case of the largest Stokes number $S = 3$, for which $a_p$ is nearly of the scale of the smallest velocity structures in the Faraday flow (compare $a_p \approx  3000 \; \mu m$ to half the Faraday wavelength $\approx  5000 \; \mu m$).} %
The approximated maximal Reynolds number of the Faraday flow is $Re_f = \frac{UL}{\nu}\approx 50$ and is calculated using the Faraday wavelength $L = \lambda_f \approx 10 \,mm$ and the root-mean-square velocity of all temporal velocity fields $ U = v_{rms} \approx 5 mm/s$ and the viscosity of water at $21$°C.

For all three flow fields, for Stokes numbers of $S=1$ and \revb{$S=3$}, the computed final positions are substantially different when ignoring the BHT.
Furthermore, the high standard deviations suggest that there are large differences between particles in the way that ignoring the BHT affects their trajectories.
It is clear that the BHT does not have the same effect on all trajectories but that there is a more complex process at play. In addition, ignoring the BHT leads to a much higher leakage of particles from the region $M$ they were initialized in, see Table \ref{tab:leakage}.
\begin{table}[t]
	\centering
	\begin{tabular}{@{}ll@{\hspace{2em}}r@{\hspace{1em}}r@{}} \toprule
	\multicolumn{4}{c}{Average relative distance and standard deviation} \\ \cmidrule(r){1-2}
	Flow field & $S$ & $R=7/9$ & $R=11/9$ \\ \midrule
	Double gyre   & $0.1$ & $0.03 \pm 0.06$  & $0.03 \pm 0.05$ \\
	               & $1$   &  $0.34 \pm 0.38$ & $0.53 \pm 0.63$ \\
                  & \revb{$3$}   &  \revb{$0.39 \pm 0.50$} & \revb{$0.57 \pm 0.69$} \\[1em]
	Bickley jet & $0.1$   & $0.04 \pm 0.23$  & $0.04 \pm 0.23$ \\
	              & $1$     & $0.28 \pm 0.56$  & $0.26 \pm 0.54$ \\
	              & \revb{$3$}     & \revb{$0.31 \pm 0.56$}  & \revb{$0.28 \pm 0.55$} \\[1em]
	Faraday flow & $0.1$  & $0.69 \pm 0.58$  & $0.75 \pm 0.62$ \\
	               & $1$   & $2.45 \pm 1.59$  & $1.99 \pm 1.49$ \\
	               & \revb{$3$}   & \revb{$2.13 \pm 1.40$}  & \revb{$1.78 \pm 1.38$}  \\ \bottomrule
	\end{tabular}
	\caption{Average relative distance~\eqref{eq:distance} (and its standard deviation) between the final positions when computing the trajectories of particles with and without Basset history term.}
	\label{tab:diff}
\end{table}

\begin{table}[t]
	\centering
	\begin{tabular}{@{}ll@{\hspace{2em}}r@{\hspace{1em}}r@{\hspace{1em}}r@{\hspace{1em}}r@{}} \toprule \multicolumn{2}{c}{} & \multicolumn{2}{c}{$R=7/9$} & \multicolumn{2}{c}{$R=11/9$} \\ \cmidrule(r){3-4} \cmidrule(r){5-6} Flow field & $S$ & With BHT & Without BHT & With BHT & Without BHT \\ \midrule
	Double gyre   & $0.1$ & $0.0$  & $0.0$ & $1.6$  & $0.0$ \\
	               & $1$   & $0.0$  & $0.0$ & $17.6$  & $39.8$ \\
                  & \revb{$3$}   & \revb{$0.0$}  & \revb{$4.4$} & \revb{$30.2$}  & \revb{$43.7$} \\[1em]
	Bickley jet & $0.1$   & $1.7$  & $1.7$ & $1.7$  & $1.7$ \\
	              & $1$     & $1.7$  & $1.6$ & $1.7$  & $1.8$ \\
	              & \revb{$3$}     & \revb{$1.7$}  & \revb{$2.2$} & \revb{$1.7$}  & \revb{$3.7$} \\[1em]
	Faraday flow & $0.1$  & $50.8$  & $56.1$ & $26.5$  & $52.1$ \\
	               & $1$   & $56.9$  & $78.6$ & $35.5$  & $64.6$ \\
	               & \revb{$3$}  & \revb{$66.0$}  & \revb{$82.9$} & \revb{$43.8$}  & \revb{$67.2$} \\ \bottomrule
	\end{tabular}
	\caption{Percentage of particles leaving the initial domain $M$.}
	\label{tab:leakage}
\end{table}
\subsubsection{Double gyre}
\begin{figure*}[th]
    \centering
    \includegraphics[scale=0.85]{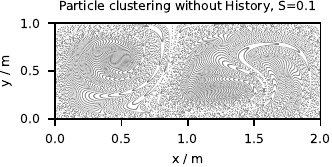}
    \includegraphics[scale=0.85]{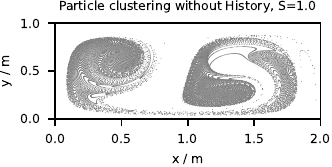}
    \includegraphics[scale=0.85]{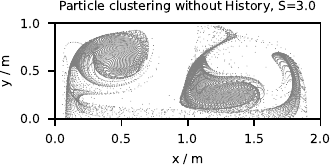} \\
    \includegraphics[scale=0.85]{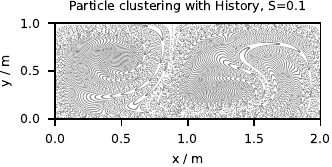}
    \includegraphics[scale=0.85]{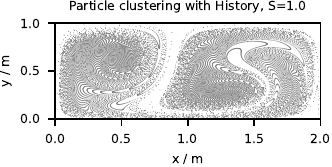}
    \includegraphics[scale=0.85]{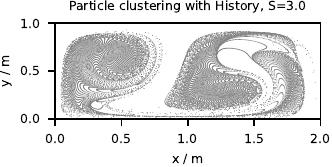}
    \caption{Particle positions confined to $M=[0,2] \times [0,1]\,\mathrm{m}^2$ at final time $t_{end}=10$ in nondimensional units, for the double gyre calculated without (top) and with (bottom) BHT. All particles have the same effective density ratio $R=\nicefrac{7}{9}$. Stokes numbers increase from left to right: $S=0.1$ (left), $S=1$ (center) and \revb{$S=3$} (right).}
    \label{fig:cluster:Gyre1}
\end{figure*}

\begin{figure*}[th]
    \centering
    \includegraphics[scale=0.85]{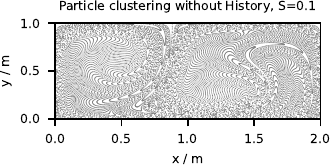}
    \includegraphics[scale=0.85]{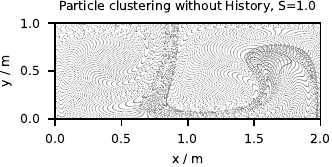}
    \includegraphics[scale=0.85]{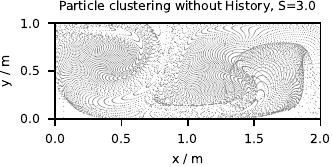} \\
    \includegraphics[scale=0.85]{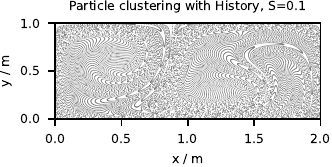}
    \includegraphics[scale=0.85]{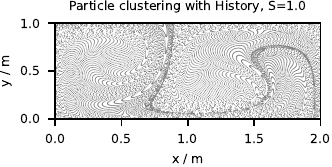}
    \includegraphics[scale=0.85]{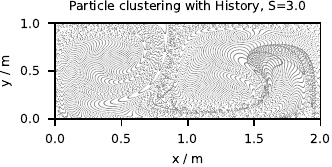}
    \caption{Particle positions confined to $M=[0,2] \times [0,1]\,\mathrm{m}^2$ at final time $t_{end}=10$ in nondimensional units, for the double gyre calculated without (top) and with (bottom) BHT. All particles have the same effective density ratio $R=\nicefrac{11}{9}$. Stokes numbers increase from left to right: $S=0.1$ (left), $S=1$ (center) and \revb{$S=3$} (right).}
    \label{fig:cluster:Gyre2}
\end{figure*}

Figure~\ref{fig:cluster:Gyre1} shows the positions of particles in the double gyre at the end of the simulation for Stokes numbers $S=0.1$ (left), $S=1$ (middle) and \revb{$S=3$} (right) for particles that are lighter than the fluid with $R=7/9$.
Figure~\ref{fig:cluster:Gyre2} shows the same for denser-than-fluid particles with $R=11/9$. 
When comparing the two particle types, it is immediately seen in all figures that lighter-than-fluid particles tend to be drawn to the vortex centers, while heavier than fluid particles are more likely to be ejected from the vortex cores~\cite{Haller2008}.
Upper figures show final positions when computing particle trajectories without history term while lower figures use the full MRE with the BHT.
Table~\ref{tab:diff} shows the average relative distance~\eqref{eq:distance} between the final position of particles computed with and without BHT.


For the lighter-than-fluid particle with $R=7/9$ (Figure \ref{fig:cluster:Gyre1}) and a small Stokes number of $S=0.1$, there is little visible impact from the history term.
However, for $S=1$ \revb{and $S=3$}, neglecting the BHT leads to much more pronounced clustering, although the patterns that form are still broadly similar.
\revb{For the case $S=3$ without BHT, we observe particles leaving the initial rectangle $M$, see Table~\ref{tab:leakage}.}
This agrees with the findings of Candelier et al.~\cite{CandelierEtAl2004}, who found that ignoring the BHT when simulating particles in a vortex flow leads to an overestimation of ejection.
This could be problematic in many cases, one example being sorting particles by Stokes number using their different separation characteristics~\cite{TallapragadaEtAl2008}.

The picture looks similar for the denser-than-fluid particle with $R=11/9$ (Figure \ref{fig:cluster:Gyre2}). For $S=0.1$, there is little difference between clustering patterns computed with and without BHT.
A visible difference emerges for $S=1$, where the inclusion of the BHT leads to slightly thinner patterns of high particle concentrations. \revb{Without BHT much }more particles are ejected from the rectangle $M$ (17.6\% of the particles with BHT vs 39.8\% without BHT, see Table~\ref{tab:leakage}.) 
For \revb{$S=3$} we also observe stronger clustering of the particles staying in $M$, when ignoring the BHT, however the resulting structures look irregular. With BHT, particle concentration is, as expected, higher along the attracting LCS (which can be related to the unstable manifold of a chaotic saddle in the open flow), even more pronounced for $S=1$ than for \revb{$S=3$}. Due to the relatively short flow time, we only see footprints of these underlying open flow structures in the particle concentrations. For \revb{$S=3$}, we also observe the expected accentuated ejection for the denser-than-fluid particles, which is again more pronounced when the BHT is ignored. 
This is also in accordance with observations that neglecting the history terms changes the number and the nature of the attractors of the system in a way that greatly affects horizontal spreading.\cite{Guseva2013}. 
\subsubsection{Bickley jet}
\begin{figure*}[th]
    \centering
    \includegraphics[scale=0.85]{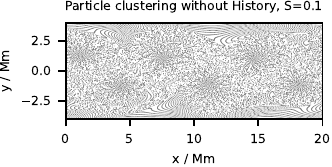}
    \includegraphics[scale=0.85]{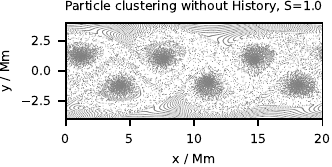}
    \includegraphics[scale=0.85]{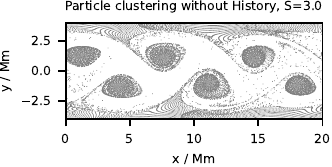} \\
    \includegraphics[scale=0.85]{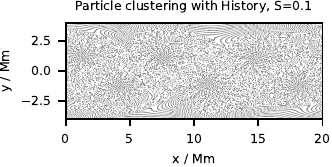}
    \includegraphics[scale=0.85]{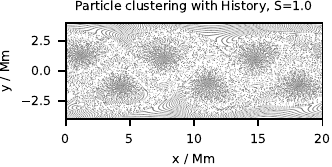}
    \includegraphics[scale=0.85]{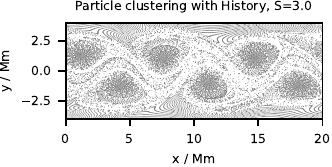}
    \caption{Particle positions confined to $M=[0, 20[ \times [-4,4]\,\mathrm{Mm}^2$ at final time $t_{end}=30$ in nondimensional units, for the Bickley jet calculated without (top) and with BHT (bottom). All particles have the same effective density ratio $R=\nicefrac{7}{9}$. Stokes numbers increase from left to right: $S=0.1$ (left), $S=1$ (center) and \revb{$S=3$} (right).}
    \label{fig:cluster:Bickley1}
\end{figure*}
\begin{figure*}[th]
    \centering
    \includegraphics[scale=0.85]{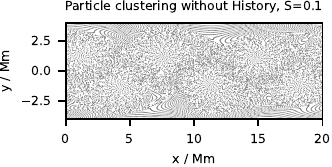}
    \includegraphics[scale=0.85]{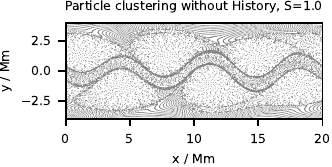}
    \includegraphics[scale=0.85]{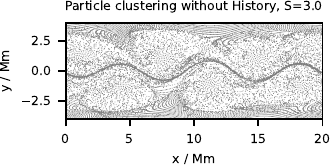} \\
    \includegraphics[scale=0.85]{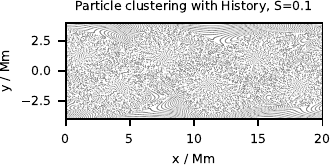}
    \includegraphics[scale=0.85]{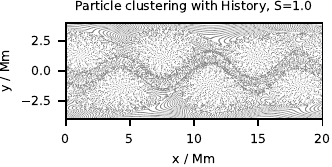}
    \includegraphics[scale=0.85]{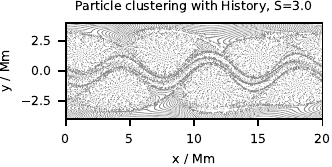}
    \caption{Particle positions confined to $M=[0, 20[ \times [-4,4]\,\mathrm{Mm}^2$ at final time $t_{end}=30$ in nondimensional units, for the Bickley jet calculated without (top) and with (bottom) BHT. All particles have the same effective density ratio $R=\nicefrac{11}{9}$. Stokes numbers increase from left to right: $S=0.1$ (left), $S=1$ (center) and \revb{$S=3$} (right).}
    \label{fig:cluster:Bickley2}
\end{figure*}
Figures~\ref{fig:cluster:Bickley1} shows the positions of particles in the Bickley jet at the end of the simulation for Stokes numbers $S=0.1$ (left), $S=1$ (middle) and \revb{$S=3$} (right) for particles with $R=7/9$ that are lighter than the fluid.
Figure~\ref{fig:cluster:Bickley2} shows the same for denser-than-fluid particles with $R=11/9$.

\begin{figure*}[h]
    \centering
    \includegraphics[scale=0.8]{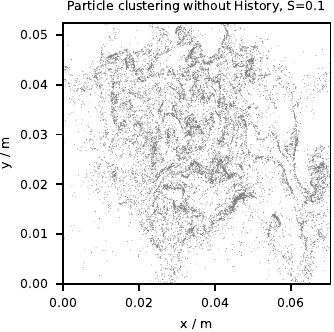}
    \includegraphics[scale=0.8]{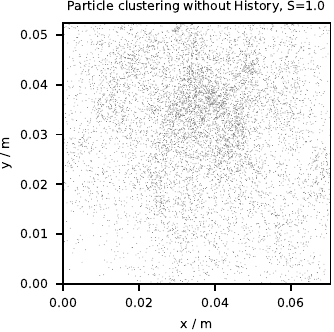}
    \includegraphics[scale=0.8]{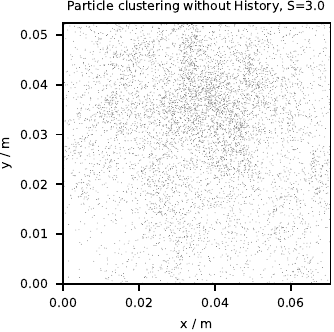} \\
    \includegraphics[scale=0.8]{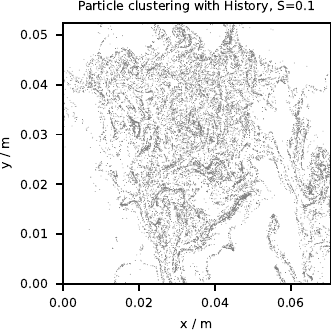}
    \includegraphics[scale=0.8]{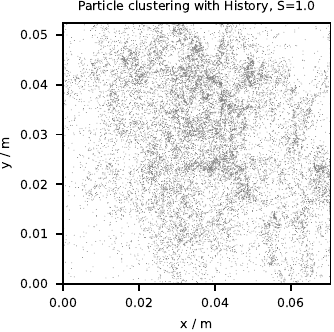}
    \includegraphics[scale=0.8]{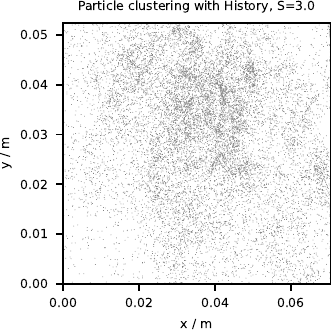}
    \caption{\reva{Particle positions calculated without (top) and with (bottom) BHT at final time, $t=\SI{10}{\second}$, for particles moving in the Faraday Flow, confined to the rectangle $M=[0,0.070395]\,m\times[0,0.052487]\,m$. All particles hold the same effective density ratio, $R=\nicefrac{7}{9}$. Stokes numbers grow from left to right: $S=0.1$ (left), $S=1$ (center) and $S=3$ (right).}}
    \label{fig:cluster:Faraday1}
\end{figure*}

\begin{figure*}[h]
    \centering
    \includegraphics[scale=0.8]{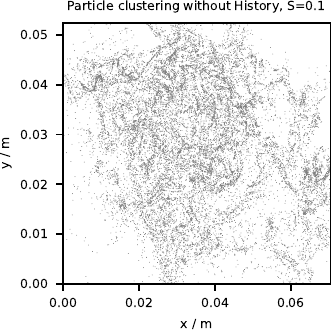}
    \includegraphics[scale=0.8]{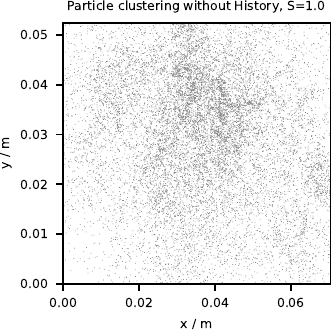}
    \includegraphics[scale=0.8]{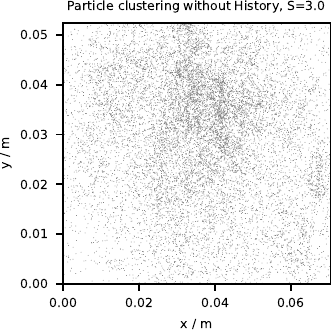} \\
    \includegraphics[scale=0.8]{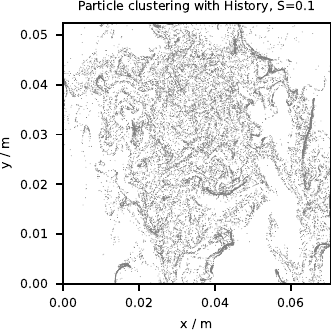}
    \includegraphics[scale=0.8]{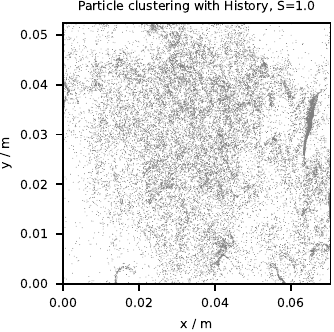}
    \includegraphics[scale=0.8]{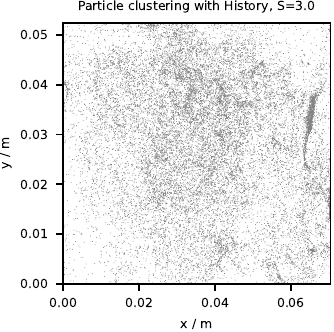}
    \caption{\reva{Particle positions calculated without (top) and with (bottom) BHT at final time, $t=\SI{10}{\second}$, for particles moving in the Faraday Flow, confined to the rectangle $M=[0,0.070395]\,m\times[0,0.052487]\,m$. All particles hold the same effective density ratio, $R=\nicefrac{11}{9}$. Stokes numbers grow from left to right: $S=0.1$ (left), $S=1$ (center) and $S=3$ (right).}}
    \label{fig:cluster:Faraday2}
\end{figure*}

As for the double gyre, the BHT has little impact for $S=0.1$, both for the lighter- and denser-than-fluid particles.
For lighter-than-fluid particles and $S=1$, neglecting the BHT leads again to stronger clustering in the centers of the vortices.
For the denser-than-fluid particles, this effect is reversed and ignoring the BHT leads to stronger ejection from the vortices.
Moreover, as shown in table \ref{tab:leakage}, there is very little leakage from the region $M=[0, 20[ \times [-4,4]\,\mathrm{Mm}^2$ the particles were initialized in, at least for $S=0.1$ and $S=1$. For \revb{$S=3$}, however, ignoring the BHT, leads to a large increase in the numbers of particles leaving $M$ both for $R=7/9$ and $R=11/9$, while the values are still small for the cases with BHT. This is in accordance with the observations made in the double gyre. The preferential particle concentration is also affected by this leakage. Again, ignoring the BHT produces more irregular patterns, especially for larger $S$. 
\reva{
\subsubsection{Faraday flow}
Figures~\ref{fig:cluster:Faraday1} shows the positions of particles in the Faraday flow at the end of the simulation using the experimentally derived velocity field for Stokes numbers $S=0.1$ (left), $S=1$ (middle) and $S=3$ (right) for particles with $R=7/9$ that are lighter than the fluid.
Figure~\ref{fig:cluster:Faraday2} shows the same for denser-than-fluid particles with $R=11/9$. 
As in the previous examples, we only plot those particles that have not left the rectangle $M=[0,0.070395]\,m\times[0,0.052487]\,m$ they were initialized in. 
Due to the large proportion of particles that do leave $M$, the ones remaining in $M$ are preferentially concentrated along fractal structures. 
This is expected in such open flows and especially visible for the small Stokes number $S=0.1$ (Figures~\ref{fig:cluster:Faraday1} and ~\ref{fig:cluster:Faraday2}, left panels). 
There are already notable differences in the final particle positions with and without BHT. 
For the larger Stokes numbers $S=1$ and $S=3$, the plots become increasingly structureless. 
This fuzziness is again more pronounced when the BHT is neglected.}

\subsection{Finite-time Lyapunov exponents}\label{sec:LCS}
This section shows colormap plots of the forward-time FTLE for all three flow fields.
For every flow field, we show two sets of nine plots, each set showing the FTLE results obtained without and with history term \revboth{(i.e.\ $\sigma_{t_0, \text{Stokes} }^{t_{end}}$ and $\sigma_{t_0, \text{History}}^{t_{end}}$, respectively, always multiplied by $\lvert t_{end}-t_0\rvert$ in the following, as described in Section \ref{sec:FTLE_background}} as well as the difference between the two respective fields in percent for Stokes numbers $S=0.1$, $S=1$, \revb{$S=3$} and density ratios $R=7/9$ and $R=11/9$ .
The relative difference between FTLEs $\Delta\sigma_{t_0,relative}^{t_{end}}$ is calculated from \revboth{
\begin{eqnarray}
    \label{eq:FTLEdiff}
    \Delta \sigma_{t_0, relative}^{t_{end}} := 100 \; \frac{\sigma_{t_0, \text{History}}^{t_{end}} - \sigma_{t_0, \text{Stokes} }^{t_{end}}}{|| \sigma_{t_0, \text{History}}^{t_{end}} ||_{l_\infty(D)}}
\end{eqnarray}
}
where $D \subset \Omega \times [t_0, t_{end}]$ and $\Omega \subseteq \mathbb{R}^2$ is the spatial domain.
The differences are signed and negative values correspond to areas where neglecting the BHT will increase FTLE while positive areas mean that neglecting the BHT will lead to smaller FTLE.

For $R=1$ the FTLE fields are independent of $S$ and, since we do not consider any effects that induce non-zero relative velocities, they agree independent of whether the BHT is present or not.
We confirmed that our numerical solutions reproduce this behavior correctly but do not show any figures.

Figures~\ref{fig:FTLE:Gyre1} to~\ref{fig:FTLE:Faraday2} below show the FTLE without history term (left), with history term (center) and the relative difference between the two (right) for different flow fields.
For the left and center plots, the colorbars indicate the value of the FTLE \revb{$\sigma_{t_0, \text{Stokes}}^{t_{end}}$ (left) and $\sigma_{t_0, \text{History}}^{t_{end}}$ (center)} at a given point $(x,y)$ (initial position of the particle), computed using equation~\eqref{eq:FTLE}. The right figures show the relative difference between the FTLE in percent, computed using~\eqref{eq:FTLEdiff}.
\subsubsection{Double gyre}
\begin{figure*}[!ht]
    \centering
    \includegraphics[scale=1]{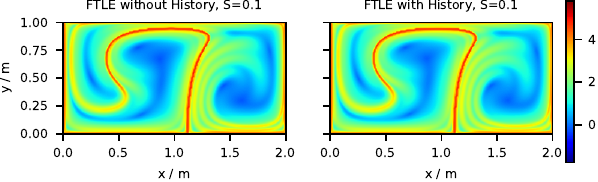}
    \includegraphics[scale=1]{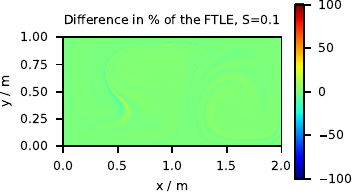} \\
    \includegraphics[scale=1]{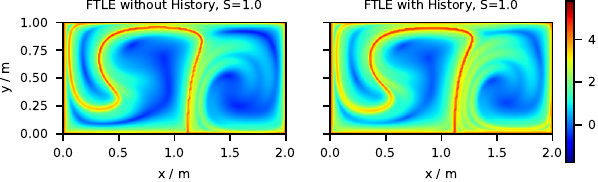}
    \includegraphics[scale=1]{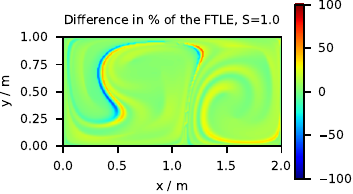} \\
    \includegraphics[scale=1]{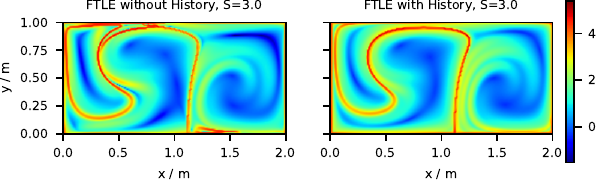}
    \includegraphics[scale=1]{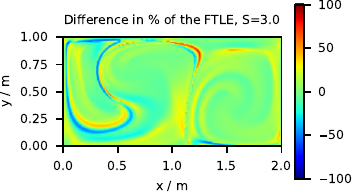} \\
    \caption{FTLE for light particles ($R=\nicefrac{7}{9}$) and three Stokes numbers, $S=0.1$ (top), $S=1$ (center), \revb{$S=3$} (bottom), being advected by the double gyre. Left and center columns show the FTLE computed with trajectories calculated without (left) and with (center) BHT. Right column shows the relative difference in the FTLE.}
    \label{fig:FTLE:Gyre1}
\end{figure*}

Figure~\ref{fig:FTLE:Gyre1} shows the FTLE without BHT (left), with BHT (center) and the difference between the two (right) for lighter-than-fluid particles with $R=7/9$.
From top to bottom the Stokes number changes from $S=0.1$ to $S=1$ and \revb{$S=3$}.

In line with the clustering plots, the impact of the BHT for $S=0.1$ is small.
The FTLE show no visible differences and agree to within a few percentage points.
For $S=1$, the FTLE with BHT still look qualitatively similar to those without BHT although some differences emerge.
In particular, the region of strong divergence between the gyres shifts it position.
When the Stokes number becomes even larger, at \revb{$S=3$}, we start to see substantial qualitative differences as well.
Without BHT, a number of filaments of strong divergence appear that are not present in the FTLE with history term.
There are now substantial differences between the two FTLE fields of up to $\pm100$\% in some regions.

\begin{figure*}[!ht]
    \centering
    \includegraphics[scale=1]{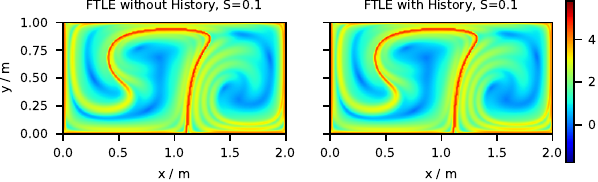}
    \includegraphics[scale=1]{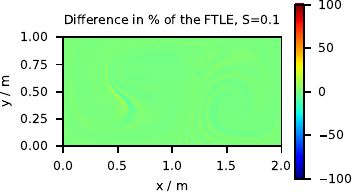} \\
    \includegraphics[scale=1]{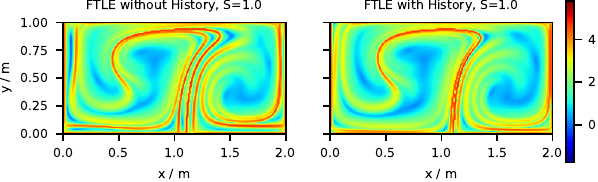}
    \includegraphics[scale=1]{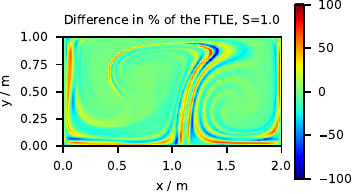} \\
    \includegraphics[scale=1]{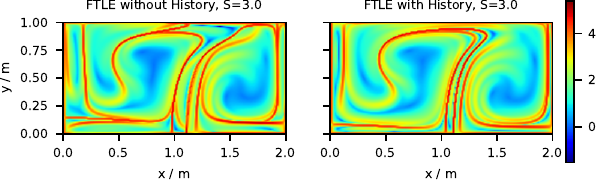}
    \includegraphics[scale=1]{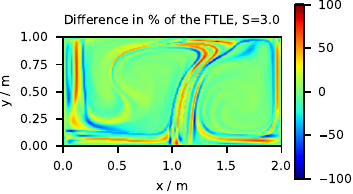} \\
    \caption{FTLE for heavy particles ($R=\nicefrac{11}{9}$) and three Stokes numbers, $S=0.1$ (top), $S=1$ (center), \revb{$S=3$} (bottom), being advected by the double gyre. Left and center columns show the FTLE computed with trajectories calculated without (left) and with (center) BHT. Right column shows the relative difference in the FTLE.}
    \label{fig:FTLE:Gyre2}
\end{figure*}

Figure~\ref{fig:FTLE:Gyre2} shows the same plots but for heavier-than-fluid particles with $R=11/9$.
As for the light particle, there is little difference between the FTLE for $S=0.1$.
For $S=1$, however, the differences between BHT and no BHT are more pronounced.
While still qualitatively similar, the regions of strong separation between and around the gyres look noticeably different.
Without BHT, there are three well separated filaments of strong divergence between the gyres which are much closer together and almost fused when the BHT is considered.
More generally, the red, high-FTLE filaments around the gyres shift away from the center of the domain when including the history term. Regions of large FTLE values highlight repelling LCS. When viewing the double gyre for inertial particles as an open or leaking systems, it seems that these repelling LCS provide an inverse picture of the preferential particle concentrations along attracting LCS in Figures \ref{fig:cluster:Gyre1} and \ref{fig:cluster:Gyre2}.  
\subsubsection{Bickley jet}
\begin{figure*}[!ht]
    \centering
    \includegraphics[scale=1]{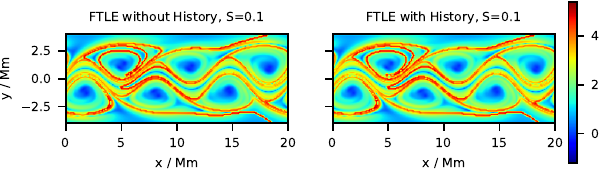}
    \includegraphics[scale=1]{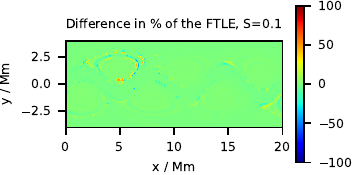} \\
    \includegraphics[scale=1]{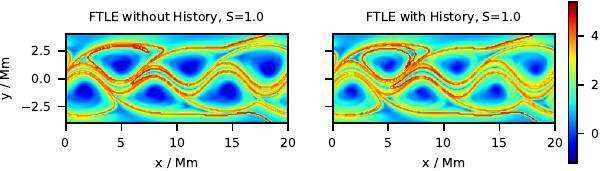}
    \includegraphics[scale=1]{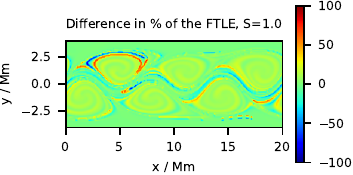} \\
    \includegraphics[scale=1]{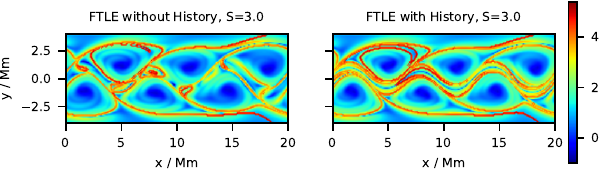}    \includegraphics[scale=1]{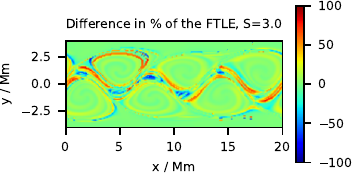} \\
    \caption{FTLE for light particles ($R=\nicefrac{7}{9}$) and three Stokes numbers, $S=0.1$ (top), $S=1$ (center), \revb{$S=3$} (bottom), being advected by the Bickley jet. Left and center columns show the FTLE computed with trajectories calculated without (left) and with (center) BHT. Right column shows the relative difference in the FTLE.}
    \label{fig:FTLE:Bickley1}
\end{figure*}

\begin{figure*}[!ht]
    \centering
    \includegraphics[scale=1]{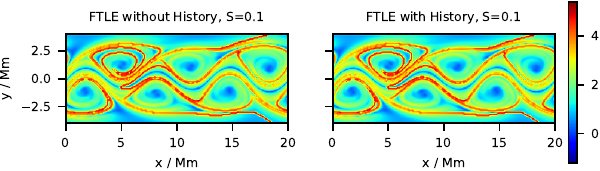}
    \includegraphics[scale=1]{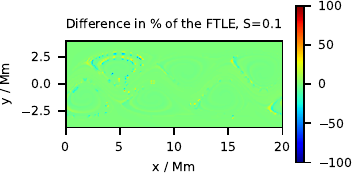} \\
    \includegraphics[scale=1]{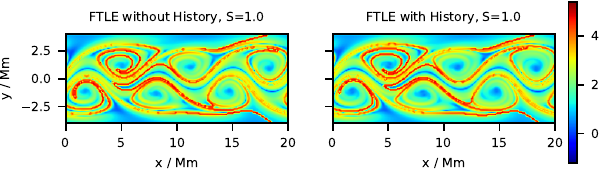}
    \includegraphics[scale=1]{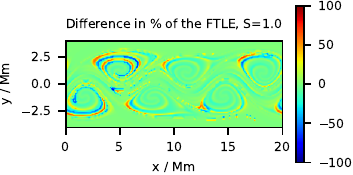} \\
    \includegraphics[scale=1]{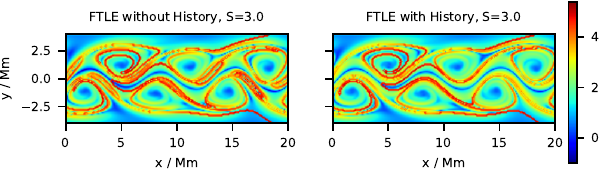}
    \includegraphics[scale=1]{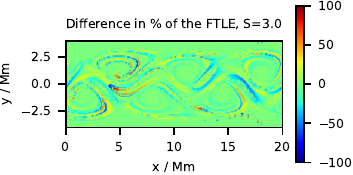} \\
    \caption{FTLE for heavy particles ($R=\nicefrac{11}{9}$) and three Stokes numbers, $S=0.1$ (top), $S=1$ (center), \revb{$S=3$} (bottom), being advected by the Bickley jet. Left and center columns show the FTLE computed with trajectories calculated without (left) and with (center) BHT. Right column shows the relative difference in the FTLE.}
    \label{fig:FTLE:Bickley2}
\end{figure*}

Figures~\ref{fig:FTLE:Bickley1} and~\ref{fig:FTLE:Bickley2} shows the FTLE (left and center) and difference between FTLE with and without BHT (right) for the Bickley jet.
Again, for both lighter- and heavier-than-fluid particles, the differences for $S=0.1$ are mostly small, although a few localized regions emerge where shifts in the position of divergence zones around the vortices lead to differences of over $50$\%.

This effect becomes more pronounced for $S=1$ where the FTLE in the jet are now showing differences up to $100$\%. Again, for \revb{$S=3$}, the differences increase and ignoring the BHT leads to a strong overestimation of FTLE of up to $100$\% in some parts.
Especially, for lighter-than-fluid particles, knot-like structures emerge in regions of high curvature within the central jet regions for the case \revb{$S=3$, for the setting without BHT (see Figure \ref{fig:FTLE:Bickley1}, bottom row, left)}. While these look like numerical artifacts on first sight, they can be well explained by the fact that particles initialized in the jet may leave it due to the inertial effects and may enter the neighborhood of one of the surrounding vortices. This is also visible in the corresponding final particle positions: compare the \revb{top right panel in Figure \ref{fig:cluster:Bickley1} where the center of the jet is almost fully depleted of particles.} 
\revb{Similar, but more elongated structures are also visible in the respective FTLE field for $R=11/9$ and $S=3$.}  
Finally, \revb{for $S=3$ and without BHT, we observe a breakdown of the jet core, both for the lighter- and denser-than-fluid particles, that is also visible in the FTLE differences (see Figures~\ref{fig:FTLE:Bickley1} and~\ref{fig:FTLE:Bickley2}, bottom panels).} 
This is a drastic change in the global dynamics and could lead to a wrong prediction of the inertial particle dynamics in fluid systems with jets and vortices as occurring in the ocean.

\subsubsection{Faraday flow}
\begin{figure*}[!htp]
    \centering
    \includegraphics[scale=1]{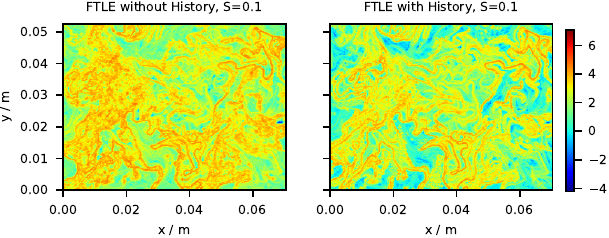}
    \includegraphics[scale=1]{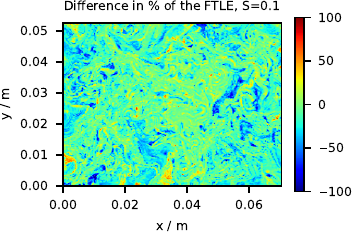} \\
    \includegraphics[scale=1]{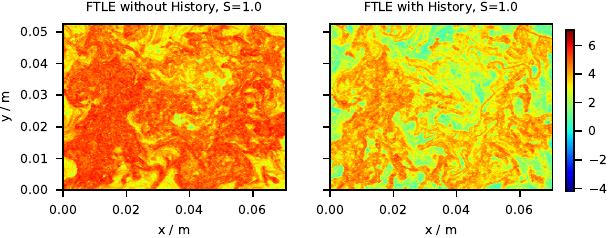}
    \includegraphics[scale=1]{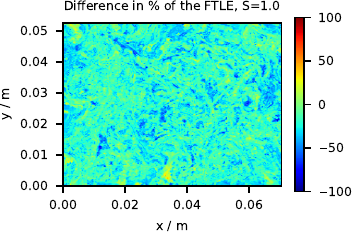} \\
    \includegraphics[scale=1]{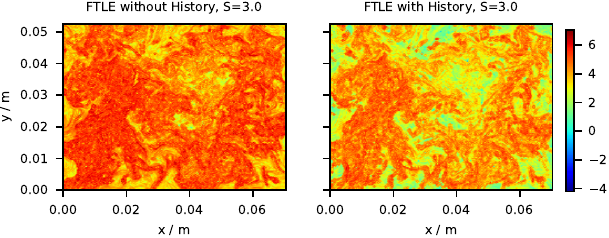}    \includegraphics[scale=1]{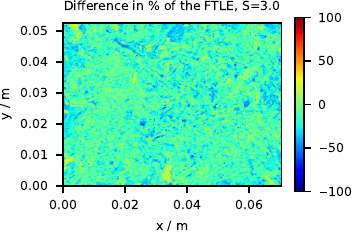} \\
    \caption{FTLE for light particles ($R=\nicefrac{7}{9}$) and three Stokes numbers, $S=0.1$ (top), $S=1$ (center), \revb{$S=3$} (bottom), being advected by the Faraday flow. Left and center columns show the FTLE computed with trajectories calculated without (left) and with (center) BHT. Right column shows the relative difference in the FTLE.}
    \label{fig:FTLE:Faraday1}
\end{figure*}

\begin{figure*}[!htp]
    \centering
    \includegraphics[scale=1]{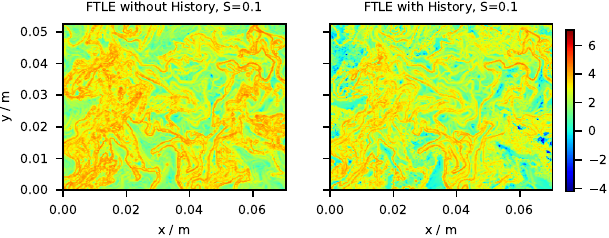}
    \includegraphics[scale=1]{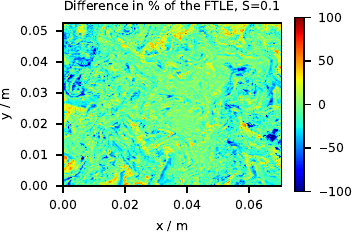} \\
    \includegraphics[scale=1]{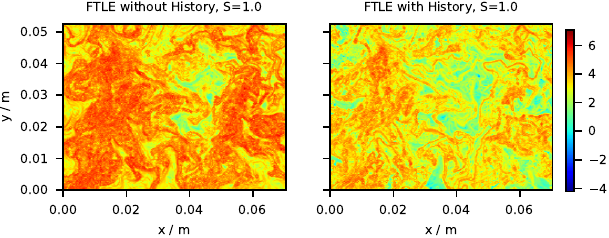}
    \includegraphics[scale=1]{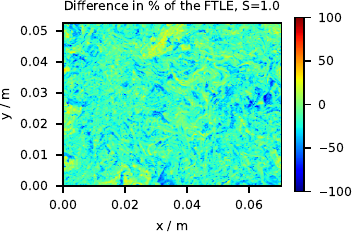} \\
    \includegraphics[scale=1]{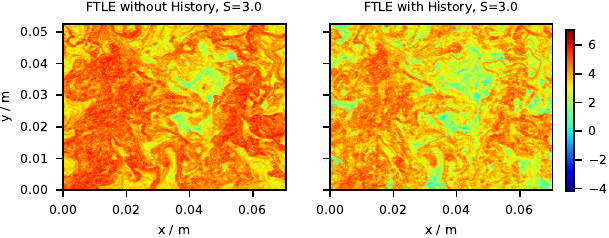}
    \includegraphics[scale=1]{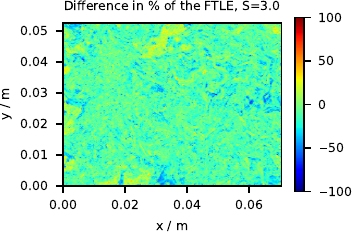} \\
    \caption{FTLE for heavy particles ($R=\nicefrac{11}{9}$) and three Stokes numbers, $S=0.1$ (top), $S=1$ (center), \revb{$S=3$} (bottom), being advected by the Faraday flow. Left and center columns show the FTLE computed with trajectories calculated without (left) and with (center) BHT. Right column shows the relative difference in the FTLE.}
    \label{fig:FTLE:Faraday2}
\end{figure*}

Finally, figures~\ref{fig:FTLE:Faraday1} and~\ref{fig:FTLE:Faraday2} show the FTLE and differences between FTLE for the Faraday flow.
For both $R=7/9$ and $R=11/9$ we see locally concentrated significant quantitative differences already for $S=0.1$.
The effect is more pronounced for the heavier-than-fluid particles where substantial regions with differences in the FTLE between $50$ and $100$\% arise.
Table~\ref{tab:diff} substantiates this observations, showing that, even on average, final particle positions are very different with and without BHT.

For $S=1$ and both lighter and heavier particles, there are much larger zones with high, positive FTLEs without BHT than with.
The effect is even stronger for \revb{$S=3$} where throughout most of the domain the FTLE without BHT are larger than $+5$ whereas with BHT significant areas with smaller FTLE of around $2-3$ remain.
In both cases, the FTLE are still locally overestimated when ignoring the BHT.
This suggests that ignoring the BHT may lead to unrealistically effective mixing in simulations, because particles trajectories diverge too rapidly. The Faraday flow can be viewed as an open system, with a large amount of particles leaving the rectangle $M$ that corresponds to the experimental field of view. However, by our simple extension of the velocity field outside of $M$, we have still obtained full-length trajectories for the FTLE computations. Regions of large FTLE values highlight initial particle positions on repelling LCS.  It would be interesting to study the relation of these flow structures of our artificially closed system to the stable manifolds of chaotic saddles for the open system \cite{Aref2017frontiers}.
\section{Conclusions}\label{sec:conclusion} %
We investigate how the Basset history term (BHT) in the Maxey-Riley equation changes the Lagrangian dynamics of simulated inertial particles.
To this end, we solve the nondimensional Maxey-Riley equation with and without BHT for thousands of particles in three flow fields, a double gyre, the Bickley jet and an experimentally measured Faraday flow.
We compare the clustering of the particles and finite-time Lyapunov exponents for trajectories computed with and without BHT.

For the double gyre and Bickley jet and a Stokes number of $S=0.1$ we see little difference between the dynamics computed with and without BHT.
Since the Faraday flow is more turbulent, even for $S=0.1$ ignoring the BHT already has a visible impact on the FTLE field.
For $S=1$ and even more so for \revb{$S=3$}, these significant difference also emerge for the double gyre and the Bickley jet.
In line with previous findings, ignoring the BHT leads to an overestimation of ejection of particles from vortices in the flow.
It also shifts the positions of areas of strong divergence, even when the overall patterns still look similar.

Generally, ignoring the BHT overestimates FTLE and leads to Lagrangian dynamics that would only be seen for a larger Stokes number if the BHT was considered.
This also makes sense mathematically: with the BHT, the full history of the particle influences the forces acting on a particle at some given time.
Without BHT, only the instantaneous forces due to the material derivative of the flow field and Stokes drag act on the particle.
Therefore, without BHT, the forces from one time step to the next will change more drastically whereas the BHT adds a form of ``mathematical inertia'' since the integral over the particle's past trajectory changes more slowly than the acting forces and thus the particle's speed and direction will change less rapidly.
This observation is in line with previous findings by Daitche~\cite{Daitche2015}, who showed that the history term causes particles to behave more like passive tracers.

In conclusion, our analysis backs up previous studies~\cite{CandelierEtAl2004,Guseva2013,prasath2019accurate,Daitche2015} that demonstrate that the Basset history term cannot safely be ignored even when simulating comparatively small particles.
It confirms that the differences matter not only at the level of individual particle trajectories but that also the larger scale Lagrangian dynamics change potentially significantly if the BHT is neglected.
In particular, the overestimation of FTLE could lead to unrealistically efficient mixing in simulations when the BHT is ignored.
Given that recent advances in numerical mathematics now allow for the efficient solution of the full MRE~\cite{Daitche2013,prasath2019accurate,urizarna2024efficient}, we argue that simulations of inertial particles should routinely include the history term especially when considering Stokes numbers $S \approx 1$ or larger. In the future we aim to develop a type of Maxey-Riley equation for larger particles that describes the movement of Lagrangian Sensor Particles in chemical reactors.


\begin{acknowledgments}
This project is funded by the Deutsche Forschungsgemeinschaft (DFG, German Research Foundation) – SFB 1615 – 503850735.
The data for the Faraday flow was obtained in project 395843083 funded by the Deutsche Forschungsgemeinschaft.
\end{acknowledgments}

\section*{Data Availability Statement}


\begin{center}
\renewcommand\arraystretch{1.2}
\begin{tabular}{| >{\raggedright\arraybackslash}p{0.3\linewidth} | >{\raggedright\arraybackslash}p{0.65\linewidth} |}
\hline
\textbf{AVAILABILITY OF DATA} & \textbf{STATEMENT OF DATA AVAILABILITY}\\  Data openly available in a public repository that issues datasets with DOIs
&
The data that support the findings of this study are openly available at \url{https://doi.org/10.5281/zenodo.14252124}, Ref.\ \onlinecite{code2024}.
\\\hline
\end{tabular}
\end{center}



\bibliography{biblio.bib}

\end{document}